# Techniques in high-speed imaging and X-ray micro-computed tomography for characterisation of iron ore fragmentation

Aleese Barron[a], Yulai Zhang, Neelima Kandula[a], Matthew Shadwell[a], Scott Bensley[a], Tim Evans[a], Louisa O'Connor[a], Mark Knackstedt[a], and Nicolas Francois[a]

[a,] *Department of Materials Physics, Research School of Physics, Australian National University, Canberra, ACT, 2601, Australia*

**ABSTRACT**

Fragmentation and breakage of rocks is essential to iron ore mining and extraction. The breakage energy requirements and resulting ore particle size and mineral distributions are key to understanding and optimising mining and processing practices. This study combines high-speed and experimental X-ray micro-CT (micro-computed tomography) imaging with 3D image analyses to study the fragmentation of iron ore particles.

Here a case study of a hematite-rich iron ore particle is used to illustrate the application and results produced by this imaging procedure. The particle was micro-CT scanned before a high-speed camera was used to image particle breakage in a custom drop weight test, capturing the dynamic processes of particle fracturing and subsequent fragmentation at a resolution of 50 microseconds. Fragments produced were collected, micro-CT scanned and analysed in three-dimension by particle shape, size, and volume.

**KEYWORDS**

Iron ore, MicroCT, High-speed imaging, dynamic crushing, comminution

## 1. INTRODUCTION

Understanding and characterising the nature of iron ore breakage and fragmentation is essential to predicting necessary energy inputs as well as resultant products from comminution processes. In order to gain insights into these processes, it is necessary to develop methods to visualise and analyse iron ore material pre-, during and post-breakage to aid in the prediction of mechanical breakage behaviour as well as the characteristics of post-breakage progeny.

This study outlines a new, custom methodology developed to image and quantify the morphology of iron ore particles and their progeny and to correlate it to dynamic breakage behaviours through the design and construction of a simple, laboratory drop test setup. The use of both microCT and high-speed imaging in this protocol provides added insights into pre- and post-breakage materials as well as breakage processes that are generally not captured during standard drop weight or breakage testing. This paper uses a case study of a single hematite-rich iron ore particle to illustrate how this methodology is applied and the results that it could produce.

## 2. MICRO-CT

2.1. Scanning facility and techniques

Individual iron ore particles were scanned at the National Laboratory for X-ray Micro Computed Tomography at the Australian National University. The CTLab is based at the Research School of Physics of the Australian National University. One of the core strengths of the CTLab is the ability to measure, characterize, and quantify 3D physical structures based on the acquisition of high-resolution (~1 μm), high-quality (high signal to noise ratio) 3D images.

Over the past 15 years, the CTLab has conducted pioneering research in the development of advanced X-ray micro-CT techniques (Sheppard et al., 2014) that have been applied to a broad range of problems encountered in geophysics (Arns et al., 2005; Saadatfar et al., 2012; Feali et al., 2013; Qajar et al., 2013; Wildenschild & Sheppard, 2013; Herring et al., 2019; Zhang et al., 2021;Francois et al., 2022) and in model granular materials (Aste et al., 2005; Francois et al., 2013; Hanifpour et al., 2014, 2015; Saadatfar et al., 2017). The work on the morphological characterisation of iron ore fragment packings reported herein grew out of this research.

Iron ore particles were first CT scanned to capture the three-dimensional morphology of both the solid and porous phases of each particle before breakage. We employed a high-resolution Heliscan system with an optimised space-filling trajectory at a resolution of 5.38 microns/voxel (Sheppard et al., 2014). Figure 1 shows a tomographic section of a hematite-rich iron ore particle.

In the following, the study of the breakage and fragmentation of this particle is presented.

2.2. 3D image analysis

The resultant CT datasets were then segmented into solid phase and macropores based on greyscale values using a converging active contours method (Shepperd et al., 2004). The volume of the particle was then determined by the number of voxels segmented in the solid phase and the shape determined by the fitting of a Legendre ellipse to determine the three axis lengths (Besterci et al., 2001). Note that this particle presents a significant level of sub-resolution microporosity (pores smaller than the voxel size of 5.38 microns). In the present study, as a first approximation, micropores have been included in the solid phase.

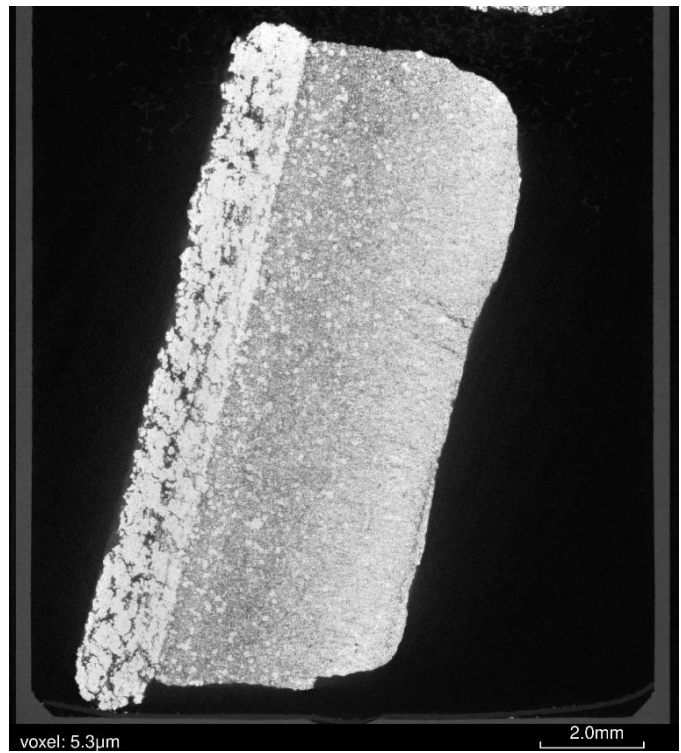

Figure 1 Tomographic cross section of the iron ore particle studied in this work.

## 3. MECHANICAL BREAKAGE TEST

3.1. Experimental setup: drop weight test

A custom drop weight test apparatus was developed to enable the capture of dynamic breakage effects using a Phantom highspeed camera (see fig. 2). This test was designed to enable the breakage of singular iron ore particles (roughly 10 x 10 x 10 mm) between a steel base and a steel impactor dropped from a 1 m height. The tested particle is accurately placed in the middle of the circular steel base. A transparent Perspex enclosure with a removable panel was placed over the steel base, which was used to contain sample breakage and capture efficiently resultant fragments while the transparent material enabled video capture with the aid of LED spotlights and a backlit screen. A Perspex platform was secured between the enclosure and the steel base to catch fragments for collection. A 20 mm hole was cut in this platform to enable fragments to be swept into a sample container placed under this hole. For each experiment, we measure the mass of the initial particle and of the fragments collected, the latter is typically larger than 90% of the initial mass.

A Perspex tube was placed vertically through the roof of the enclosure to guide the steel impacter onto the sample in the middle of the steel base. The inside of the tube was coated with a Teflon spray to reduce friction between the impacter and the tube. Perspex columns and an upper Perspex platform were affixed on top of the enclosure to ensure accurate positioning and stability of the guiding tube.

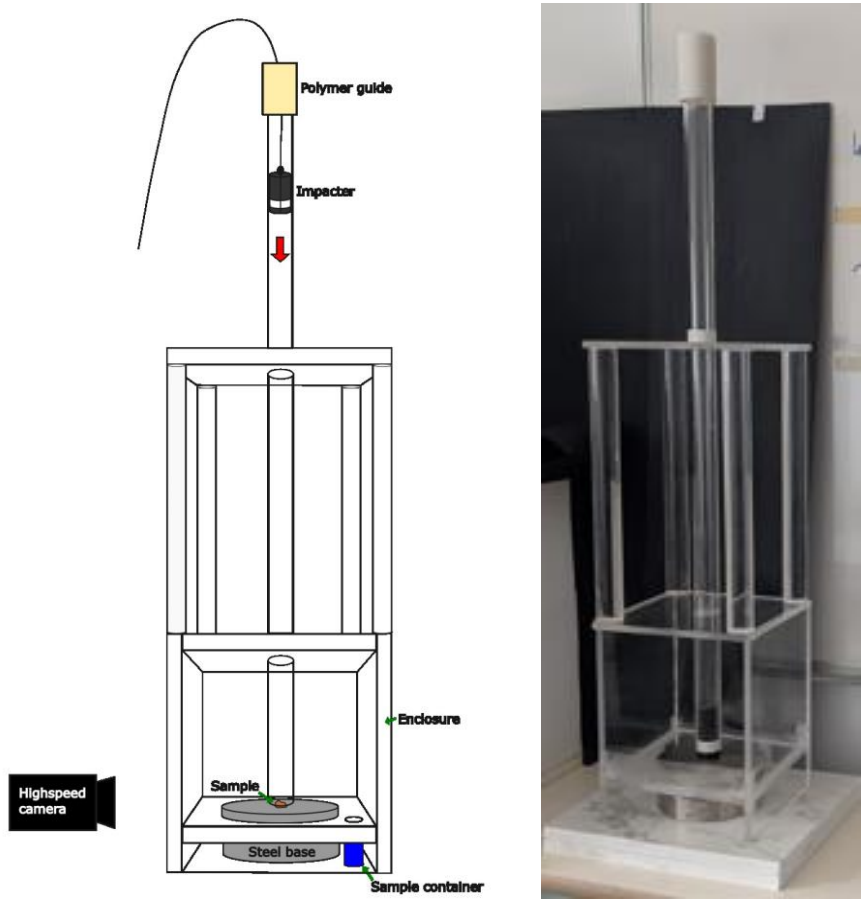

Figure 2 Left: Schematic drawing of custom drop weight test. Right: Photo of lab setup of custom drop weight test.

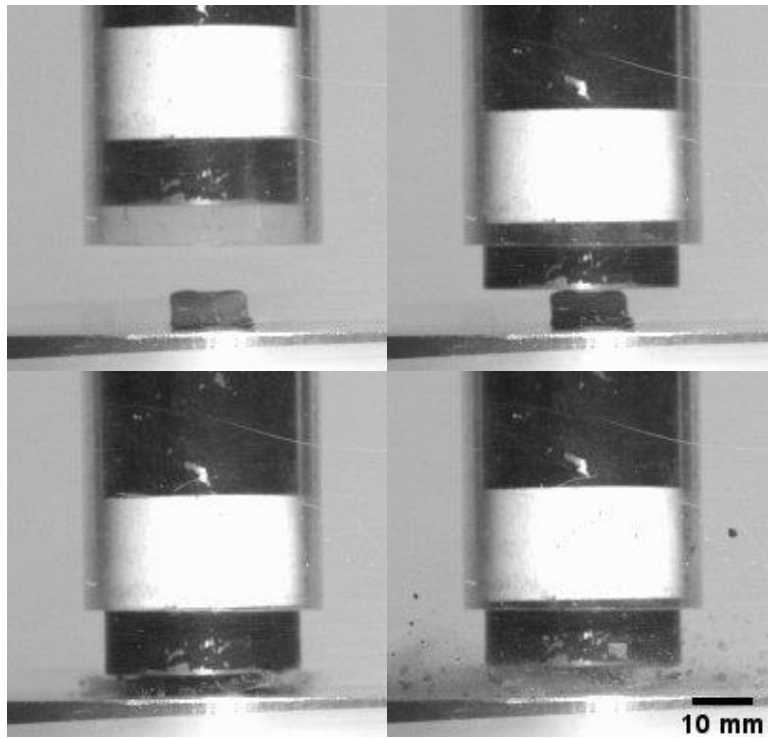

Figure 3 Single frames from highspeed video of breakage of iron ore particle. Top left: Before breakage. Top right: Frame of contact between impacter and particle. Bottom left: Frame of minimum height of impacter above base, where velocity reverses to rebound. Bottom right: Height of impacter rebound and expulsion of particle fragments.

An 84 mm length of 10.5 mm diameter steel rod with a mass of $m$ =500g was used to act as the impacter. It was painted black and a strip of white tape was added around its width to enable easy and accurate tracking during video analysis. A short length of poly mer tubing that fit snugly around the impacter was attached to the top of the Perspex tube to help centre the impacter on release. The impacter was dropped from a height of 1.05 m.

A high-speed Phantom camera was used to capture videos of both the impacter motion and of the particle breakage (see Fig. 3). Movies were acquired at 20 000 fps (50 microseconds) which gives enough temporal resolution to observe the breakage dynamics that typically occurs over time intervals of 2 milliseconds.

3.2. Image Analysis

The dynamics of the impacter were analysed by tracking the white tape on it in the highspeed videos. ImageJ software was used to binarise the images in order to highlight the tape and then track its centre of mass through each frame. The main result of this procedure is the reconstruction of the height trajectory *h(t)* of the impacter. This allowed us to compute the velocity $v(t) = dh/dt$ and acceleration $a(t) = dv/dt$ of the impactor during the breakage process.

The velocity could be accurately determined $vo = v(t = 0)$ just before impact and the kinetic energy $mvo^2/2$ was almost identical to the potential energy $mgho$ where $ho = 1.05m$ and $m = 500g$ is the mass of the impacter. This confirmed that losses by friction in the guiding tube are negligible.

The inertial force of the impactor was determined as: $F(t) = m * a(t)$. The following describes the behaviour of this force during the particle breakage. The work done by this force on the particle can be considered as the "dynamic energy input" during the breakage.

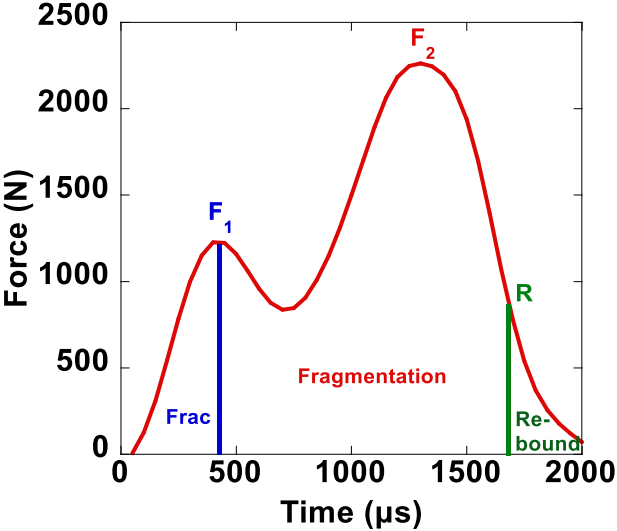

Figure 4 Impacter inertial force during breakage showing the two stages of breakage: fracturing and fragmentation.

3.3. Breakage Dynamics

Figure 4 shows the evolution of the inertial force during the breakage of the particle (particle shown in Figure 1). The time evolution of the force exhibited two peaks over a 2ms time period. Each peak corresponded to different breakage regimes.
Upon comparison with the highspeed video frames, the first peak ($F_1$), reaching approximately 1200 N, coincided with the initial fracturing of the particle into two main large fragments, this occurred 450 microseconds after the impacter made contact with the particle. This peak was followed by a brief reduction in force before a second higher peak ($F_2$) of approximately 2300 N. This second regime coincided with an obvious fragmentation of the particle. This crushing resulted in the formation of numerous fragments which were radially expelled from the gap space between the impacter and the steel base. Following this crushing phase after 1650 microseconds of contact with the particle, the velocity of the impacter reversed from negative to positive indicating a small fraction (~5%) of the input energy being converted into a rebound momentum.

The high-speed camera enabled observation of the importance of the ore fragments inertia in the crushing/fragmentation phase. Indeed, the large fragments produced during the first fracturing regime did not gain sufficient radial momentum to be expelled from the gap space formed by the impacter and the steel base. As the impacter kept moving down, this "inertial trapping" led to an efficient crushing of these large fragments into a progeny of much smaller particles.

The original single particle (Fig 1) produced more than 10,000 fragments.

## 4. TOMOGRAPHY-BASED FRAGMENT ANALYSIS

### 4.1. Collection & Imaging

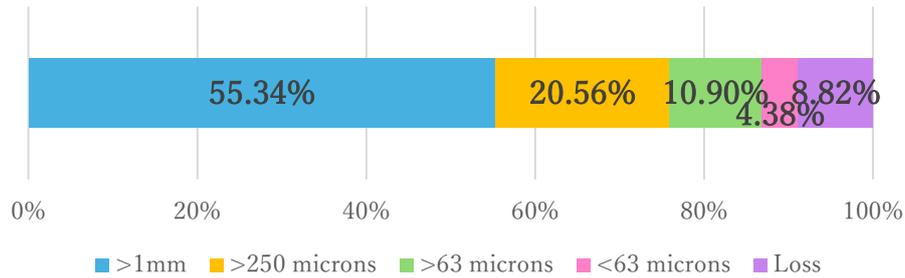

Figure 5 Sieved size fractions as percentage of original particle by mass.

After particle breakage, the resultant fragments were collected by brushing into the sample container below the hole in the Perspex platform (see figure 2). The fragments were then sieved into four size fractions (>1mm, >250 microns, >63 microns and <63 microns and weighed, see fig. 5). The larger size fractions of >1mm and >250 microns were packed into containers and CT scanned separately at voxel sizes of 3.18 microns and 2.29 microns.

### 4.2. Segmentation & labelling

Using converging active contours method, the fragment datasets were segmented into solid phase and macroporosity. Then, each fragment within the packing was identified using a modified watershed mapping method adapted for complex fragments morphologies following Zhang et al. (2024a; 2024b) and assigned a unique label (see fig. 6). This enabled the number of fragments in each size fraction to be determined precisely. For this case study, 151 fragments within the >1mm sieved fraction and 2131 fragments within the >250 microns sieved fraction were measured. Figure 7 shows high-accuracy 3D measurements of several size and shape descriptors of the fragments.

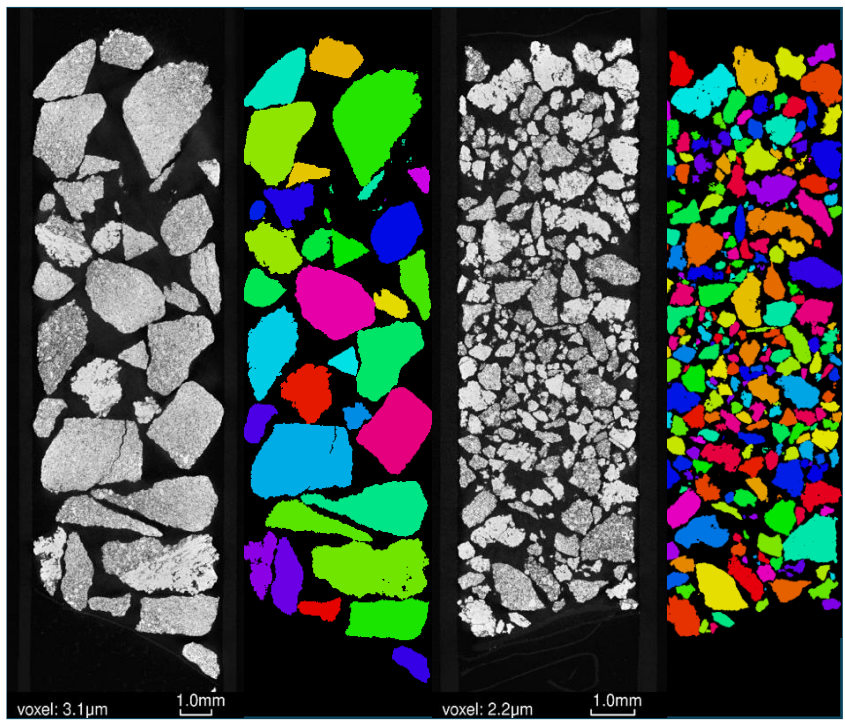

Figure 6 Tomographic slices and labelled fragments. Left: Fragments >1mm. Right: Fragments >250 microns.

Since the fragment shapes are highly anisotropic, the definition of a "particle size" is not trivial. As a first approximation, the common definition of particle size as the diameter of a sphere having the same volume as the volume of a given anisotropic particle was utilised. The initial particle had a size of 16 x 9 x 7 mm. Figure 7 (top figures) shows the broad distribution of fragment size and volumes produced by this classic drop weight test. The size distribution can be approximated by an exponential law over the range [0.2 - 2] mm.

To push further the analysis of fragments shape and describe their anisotropy, a Legendre ellipsoid was fitted to each individual fragment. The 3 main axis of such ellipsoid were used to determine a fragment length, height and width. These lengths were then plotted as length ratios on a Zingg diagram to visualise the predominant particle shapes (see. fig.7). Indeed, the Zingg diagram is an efficient tool to define 4 different shape classes, namely: spherical, disc shaped, rod-like or bladed. These diagrams highlight the broad distribution of shapes of the iron ore fragments. This variability reflects the coupling of the breakage mechanics with the heterogeneous porous structure of the initial particle.

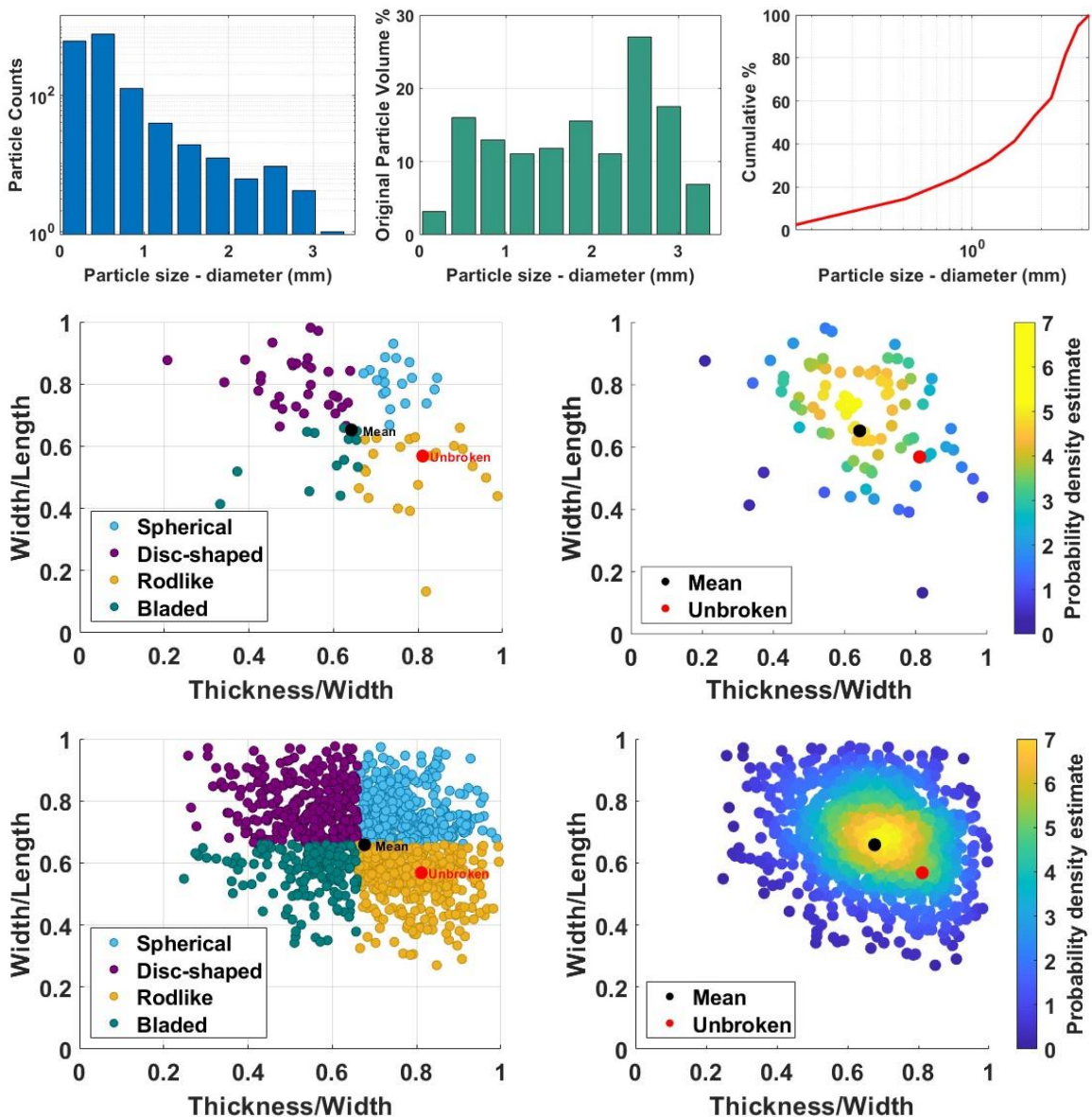

Figure 7 Fragment size and shape analysis. Top (left to right): Particle size by particle counts, particle size by volume, probability density function of particle size. Middle: Particle shape of fragments >1mm. Bottom: Particle shape of fragments >250 microns.

## 5. CONCLUSIONS

A dynamic drop weight test and imaging protocol was established to investigate the mechanical behaviour of iron ore particles during breakage as well as the size and shape of resultant fragment progeny. A laboratory drop test was constructed to capture the dynamic breakage effects occurring over few milliseconds using highspeed imaging. In this case study, analysis of the breakage of a porous hematite-rich particle revealed a two-stage breakage regime of fracturing followed by fragmentation. The created fragments were micro-CT scanned, segmented and analysed in 3D which enabled a high-accuracy description of particle size and shape. Such high-fidelity 3D information could be valuable to mineral processing and ore beneficiation.

**ACKNOWLEDGEMENTS**

N.F. acknowledges support from the Australian Research Council's Mid-Career Industry Fellowship (IM230100157).